\documentclass[12pt]{article}
\pdfoutput=1
\usepackage{latexsym, graphicx,cite} 
\usepackage{amsmath}
\usepackage{amssymb}
\usepackage[top = 1. in,bottom = 1. in, left = 1 in, right=1 in]{geometry}

 \newcommand{\arXiv}[1]{\href{http://www.arXiv.org/abs/#1}{#1}}
\usepackage[colorlinks=true, linkcolor=blue, bookmarks=true]{hyperref}

\makeatletter
\renewcommand\section{\@startsection {section}{1}{\z@}%
                                   {-3.5ex \@plus -1ex \@minus -.2ex}
                                   {2.3ex \@plus.2ex}%
                                   {\normalfont\large\bfseries}}
\renewcommand\subsection{\@startsection{subsection}{2}{\z@}%
                                     {-3.25ex\@plus -1ex \@minus -.2ex}%
                                     {1.5ex \@plus .2ex}%
                                     {\normalfont\bfseries}}
\makeatother

\newcommand{\beq}{\begin{equation}}
\newcommand{\eeq}{\end{equation}}
\newcommand{\ber}{\begin{array}}
\newcommand{\eer}{\end{array}}
\newcommand{\del}{\partial}
\newcommand{\sssty}{\scriptscriptstyle}

\newcommand{\dsty}{\displaystyle}

\newcommand{\de}{\delta}

\newcommand{\eps}{\varepsilon}

\newcommand{\om}{\omega}

\newcommand{\ena}{\end{eqnarray}}
\newcommand{\beqa}{\begin{eqnarray}}
\newcommand{\eeqa}{\end{eqnarray}}
\newcommand{\bea}{\begin{eqnarray}}
\newcommand{\eea}{\end{eqnarray}}

\begin{document}
\begin{titlepage}
\begin{flushright}
\phantom{arXiv:yymm.nnnn}
\end{flushright}
\vfill
\begin{center}
{\LARGE\bf Detailed ultraviolet asymptotics\vspace{2mm}\\for AdS scalar field perturbations}   \\
\vskip 15mm
{\large Oleg Evnin$^{a,b}$ and Puttarak Jai-akson$^{a}$}
\vskip 7mm
{\em $^a$ Department of Physics, Faculty of Science, Chulalongkorn University,\\
Thanon Phayathai, Pathumwan, Bangkok 10330, Thailand}
\vskip 3mm
{\em $^b$ Theoretische Natuurkunde, Vrije Universiteit Brussel and\\
The International Solvay Institutes\\ Pleinlaan 2, B-1050 Brussels, Belgium}

\vskip 3mm
{\small\noindent  {\tt oleg.evnin@gmail.com, puttarak.jaiakson@gmail.com}}
\vskip 10mm
\end{center}
\vfill

\begin{center}
{\bf ABSTRACT}\vspace{3mm}
\end{center}

We present a range of methods suitable for accurate evaluation of the leading asymptotics for integrals of products of Jacobi polynomials in limits when the degrees of some or all polynomials inside the integral become large. The structures in question have recently emerged in the context of effective descriptions of small amplitude perturbations in anti-de Sitter (AdS) spacetime. The limit of high degree polynomials corresponds in this situation to effective interactions involving extreme short-wavelength modes, whose dynamics is crucial for the turbulent instabilities that determine the ultimate fate of small AdS perturbations. We explicitly apply the relevant asymptotic techniques to the case of a self-interacting probe scalar field in AdS and extract a detailed form of the leading large degree behavior, including closed form analytic expressions for the numerical coefficients appearing in the asymptotics.

\vfill

\end{titlepage}


\section{Introduction}

Dynamics of small amplitude perturbations of anti-de Sitter (AdS) spacetimes has been a subject of active investigations since the intriguing observation, made in \cite{BR}, that some initial perturbations appear to collapse to small black holes no matter how much one decreases the initial perturbation amplitude. A key theme in the subsequent work on the subject (see \cite{review} for a brief review and further references) was to understand the nature of the turbulent transfer of the initial perturbation energy to shorter and shorter wavelengths, which must precede black hole formation. Questions of central importance thus lie in the ultraviolet dynamics.

The global AdS spacetime acts like a cavity with localized wave perturbations scattering, getting refocused and scattering again, each time changing their shape (only slightly in the weakly nonlinear regime). Importantly, the cavity is fully resonant with all the linearized normal mode frequencies being integer in appropriate units, which underlies the strong effect arbitrarily small nonlinearities may produce over large times.

Dynamics in the AdS `cavity' can be understood in terms of motion on different time scales. First, there are `fast' defocusing-refocusing bounces of the wave profile (for a spherically symmetric pulse of a massless field, the bounces are simply between the origin and the AdS boundary). Second, there is a `slow' energy transfer between linearized AdS normal modes due to nonlinearities in the problem. (The nonlinear terms are relatively insignificant in magnitude in the small amplitude regime and take a long time to have any appreciable effect. It is nonetheless precisely this effect that induces phenomena of physical interest.)

It is natural to attempt `integrating out' the fast motion of wave fronts and obtaining an effective theory for the slow changes of amplitudes and phases of the linearized normal modes under the influence of small nonlinearities. This approach is well-known in nonlinear physics under the name of `time-averaging' and it allows an accurate description of phenomena which are completely out of reach for naive perturbative expansions in powers of the perturbation amplitude. Time-averaging was brought up (in a slightly different guise) in relation to the problem of nonlinear gravitational stability of the AdS spacetime in \cite{BBGLL} and was given a fully analytic formulation in \cite{CEV1,CEV2}. It was subsequently applied as an alternative approach to numerical simulations (in regimes unreachable for ordinary numerical general relativity) in \cite{BMR, GMLL}, yielding results of qualitative significance. A much simpler case of a self-interacting scalar field has been given attention in \cite{BKS1,BKS2}. In that case, one expects that the turbulent energy transfer eventually saturates. However, the time-averaged approximation for small amplitude perturbations has strong similarities to the full gravitational case, and provides a useful testing ground for development of analytic techniques. (Recently, a numerical study combining gravitational nonlinearities and matter self-interactions appeared in \cite{CJY}.)

The equations of the time-averaged effective theory (see, for instance, \cite{review,CEV2}) express the first time derivatives of amplitudes and phases of the AdS normal mode oscillations as complicated nonlinear expressions in terms of amplitudes and phases themselves. The coefficients in these nonlinear expressions (which we call the \emph{interaction coefficients}) are evaluated as integrals of products of AdS mode functions and their derivatives (the precise structure appearing in these integrals is determined by the type of nonlinearities one introduces to the problem). The AdS mode functions can be expressed through Jacobi polynomials, whose algebraic features have been extensively studied. A number of special properties arise in this context, such as selection rules that enforce vanishing of large families of the interaction coefficients \cite{CEV1,CEV2,Yang,EK,EN}.

Our main objective for the rest of the paper will be to develop methods for analysis of the interaction coefficients in limits when some or all of the modes involved have extremely short wave lengths, and demonstrate how these methods work for the case of a self-interacting scalar field. Some attention has previously been given to such ultraviolet asymptotics problems in the literature \cite{CEV3,CEJV,MV}, but only with partial results for the interaction coefficients. (For example, the expressions in \cite{CEV3} only give the leading power scaling, without any access to its numerical coefficient, whereas the expressions of \cite{CEJV} only apply to the 4-dimensional AdS.) The problem relates to properties of high-degree asymptotics of Jacobi polynomials, for which a number of results are known \cite{DLMF,Elliott,Hahn,BG,WZ,int}, but applying them to the questions we are facing involves a number of subtleties. For example, subleading corrections in the asymptotics of each individual Jacobi polynomial might contribute to the leading asymptotics of an integral of their product, and different methods are required to efficiently handle the leading asymptotics depending on whether all or not all the modes involved in a particular interaction coefficient are driven to the ultraviolet regime. Our end result will be a series of expressions for the leading asymptotics of the interaction coefficients (including the numerical factors) for ultraviolet limits taken in different directions in the mode space.

We expect that our results will form a useful ingredient for subsequent analysis of weakly turbulent AdS instabilities. In particular, the self-interacting scalar field formulas we derive explicitly may contribute to proofs that the turbulent cascade saturates in this case, whereas the gravitational analog of our derivations would have implications for the more intriguing case of gravitational instability and collapse. Accurate asymptotic formulas may furthermore be utilized to drastically simplify evaluation of interaction coefficients in numerical work revolving around time-averaged AdS perturbation theory. We do not aim at pursuing these directions here and focus on the non-trivial problem of accurately recovering the ultraviolet behavior of the mode couplings.


\section{Interaction coefficients and Jacobi polynomials}

Construction of time-averaged equations for the AdS stability problem has been reviewed a number of times in the literature (see, e.g., \cite{CEJV}), and we shall only briefly quote the main formulas here to set the stage.

We consider the global AdS$_{d+1}$ spacetime with $d$ spatial dimensions. The metric (with the AdS radius set to 1) can be written as
\begin{equation}
ds^{2}=\frac{1}{\cos^2 \xi}\left(-dt^{2}+d\xi^{2}+\sin^2 \xi\,d\Omega_{d-1}^{2}\right).
\end{equation}
We have renamed the radial coordinate (usually denoted $x$ in the literature) to $\xi$ in order to reserve the letter $x$ for future use in polynomial expressions. This coordinate runs between 0 and $\pi/2$, the latter value corresponding to the AdS boundary. $d\Omega_{d-1}^{2}$ denotes the standard line element of a $(d-1)$-sphere.
\newpage
The equations of motion for a spherically symmetric\footnote{We only treat the spherically symmetric sector in this paper. For a probe scalar field, extending perturbation theory to non-spherically symmetric processes is quite straightforward. Some analysis of the resulting structures has been given in \cite{Yang,EK,EN}. On the other hand, introducing non-spherically symmetric perturbations to gravitational theories is forbiddingly complicated. Existing considerations can be consulted in \cite{nonsymm1,nonsymm2,nonsymm3}.}  massless probe scalar field $\phi(x,t)$ with a $\phi^4/4!$ nonlinearity in the action take the form
\beq
-\del_t^2 \phi +\frac{1}{\tan^{d-1}\xi}\del_\xi\left(\tan^{d-1}\xi\,\del_\xi\phi\right)=\frac{\phi^3}{3!\cos^2 \xi}.
\label{NLKG}
\eeq
The mode functions of the linearized problem (with the right-hand side above neglected) are explicitly known and given (for $n\ge 0$) by
\beq
 e_n(\xi) = k_n \,\cos^d \xi\,\,P^{(\frac{d}{2}-1,\frac{d}{2})}_n(\cos 2\xi)  \quad\mbox{with}\qquad
 k_n = 2\frac{\sqrt{n!(n+d-1)!}}{\Gamma(n+\frac{d}{2})}.
\label{modes}
\eeq
Here, $P^{(\frac{d}{2}-1,\frac{d}{2})}_n(x)$ are Jacobi polynomials of degree $n$, a family of polynomials defined to be orthogonal on the interval $[-1,1]$ with respect to the measure
\beq
\mu(x)=(1-x)^{\frac{d}{2}-1}(1+x)^{\frac{d}{2}}. 
\eeq
The frequencies corresponding to normal modes (\ref{modes}) are known to be all integer and given by
\beq
\om_n=d+2n.
\eeq
For future reference, we provide an asymptotic formula for $k_n$ at large $n$, which can be derived using Stirling's approximation:
\beq
k_n\approx \frac{2}{\sqrt{e}}\frac{n^{n/2}(n+d-1)^{(n+d-1)/2}}{(n+d/2-1)^{n+d/2-1}}\approx 2\sqrt{\frac{n}{e}} \frac{\left(1+\frac{d-1}{n}\right)^{n/2}}{\left(1+\frac{d/2-1}{n}\right)^n}\approx 2\sqrt{n}.
\label{klarge}
\eeq
(One has to be careful keeping track of the factors of $e$.) It is convenient to expand the full nonlinear equation (\ref{NLKG}) in the linearized modes by writing $\phi(x,t)=\sum_{n}c_{n}(t)e_{n}(x)$, which yields
\beq
\ddot c_n +\om_n^2 c_n =\frac{1}{3!} \sum\limits_{jkl} C_{njkl} c_jc_kc_l.
\label{eqforc}
\eeq
The above expression intoduces the \emph{interaction coefficients}
\begin{equation} \label{eq:C1}
C_{jklm} = \displaystyle{\int_0^{{\pi}/{2}}} d\xi\, \frac{\tan^{d-1}\xi}{\cos^2\xi}\, e_j(\xi)e_k(\xi)e_l(\xi)e_m(\xi),
\end{equation}
which will form the main subject of our subsequent study.

In practical applications to the AdS stability problem, one considers (\ref{eqforc}) in the small amplitude regime $c_n=O(\eps)$. On time scales of order $1/\eps^2$, (\ref{eqforc}) can be replaced by a simplified `time-averaged' equation via a well-controlled procedure known to provide an arbitrarily accurate approximation for sufficiently small $\eps$, see \cite{CEJV} for further details. (Most of the interesting phenomena discussed in relation to AdS stability happen precisely on time scales of order $1/\eps^2$.) When converting (\ref{eqforc}) to the time-averaged equations, most of the interaction coefficients $C$ drop out, except for those that satisfy the resonance relation $\om_j+\om_k=\om_l+\om_m$, i.e., $j+k=m+n$. (Other resonance conditions, such as $\om_j=\om_k-\om_l-\om_m$, for instance, might have contributed in principle, but the corresponding $C$-coefficients vanish due to AdS-specific selection rules \cite{CEV1,CEV2,Yang,EK,EN}.)

By changing variables, we can write (\ref{eq:C1}) in a polynomial form:
\begin{equation}\label{eq:C2}
C_{jklm} = \frac{k_jk_kk_lk_m}{2^{2d}}\int_{-1}^{1} dx\,\mu(x)(1+x)^{d-1}P^{(\frac{d}{2}-1,\frac{d}{2})}_j(x)P^{(\frac{d}{2}-1,\frac{d}{2})}_k(x)P^{(\frac{d}{2}-1,\frac{d}{2})}_l(x)P^{(\frac{d}{2}-1,\frac{d}{2})}_m(x).
\end{equation}
We shall focus on limits of (\ref{eq:C2}) with two, three or four mode numbers becoming large. (Sending only one mode number to infinity while keeping the other three fixed trivially produces a zero result by the Jacobi polynomial orthogonality.) All of these limits may play a role in the development of turbulent cascades in the AdS stability problem, as they describe couplings of ultraviolet modes among themselves, as well as to low-lying modes. Additionally, the interaction coefficients with three large indices play a decisive role \cite{CEJV} in determining properties of quasiperiodic solutions of the sort considered in \cite{BBGLL,GMLL}, which are dominated by a single low-lying mode.


\section{Analytic techniques for Jacobi polynomial asymptotics}

We intend to derive asymptotic representations for the interaction coefficients given by (\ref{eq:C2}) in ultraviolet limits when some or all of the indices $(j,k,l,m)$ become large. We shall start with a broad review of possible strategies one could employ for recovering this asymptotics, explaining the limitations of these techniques and highlighting the subtlety of the problem. We shall see in the end that substantially different approaches turn out to be useful for analyzing different limits of interest.

Perhaps the most natural first attempt at approximating (\ref{eq:C2}) would be to substitute an explicit representation for the Jacobi polynomials and perform the integration, followed by an attempt to extract the relevant asymptotics. A particularly attractive representation is
\beq
P^{(\alpha,\beta)}_n(x)=\sum\limits_{s=0}^n {n+\alpha \choose s}{n+\beta \choose n-s}\left(\frac{x-1}2\right)^{n-s}\left(\frac{x+1}2\right)^{s}, 
\label{Jacobigen}
\eeq
with 
\beq
{z\choose n}\equiv \frac{\Gamma(z+1)}{\Gamma(n+1)\Gamma(z-n+1)}.
\label{binom}
\eeq
Substituting this expression in (\ref{eq:C2}) results in a quadruple sum (one for each Jacobi polynomial involved) of simple integrals of powers of $(1-x)$ and $(1+x)$, where each term can be integrated explicitly in terms of Euler's beta-functions. The reason why this naive approach is not effective for extracting the large degree asymptotics is that many individual terms in this quadruple sum grow much faster as the degree of the polynomials increases than the total expression. Extracting the leading asymptotics from this sum would thus require taking into account a large (effectively infinite, in the limit of high polynomial degrees) number of cancellations between different terms, making this `brute force' approach impractical.

Compelled to turn to more sophisticated techniques for asymptotic evaluation, we recall that a number of asymptotic formulas for individual Jacobi polynomials (rather than integrals of their products) is available in the literature. A small but useful compendium can be found in \cite{DLMF}, and includes two different types of formulas: ones expressing the asymptotics through elementary (trigonometric) functions, and those expressing the asymptotics through Bessel functions. What we need to discuss is how such asymptotic expressions can be used in the context of integrals of the form (\ref{eq:C2}), rather than in the context of approximating large degree Jacobi polynomials at one point.

The trigonometric-type formulas for the Jacobi polynomial asymptotics (see the beginning of \cite{DLMF} and the original papers \cite{Elliott,Hahn}) have the advantage of presenting explicit infinite series in powers of the inverse degree of the Jacobi polynomial, with all the terms in the series having closed form expressions. Even more importantly in our context, trigonomeric functions (or algebraic functions in some complex plane versions of the same formula) are easy to integrate. A big disadvantage is that the approximation is not uniform on the interval $[-1,1]$ whereas its extensions in the complex plane require an introduction of obstructive cuts. In \cite{CEJV}, this approach was used to develop asymptotic expressions relevant for AdS$_4$, where the non-uniformity of convergence appears to have no effect on the accuracy of the approximation in (\ref{eq:C2}). On the other hand, in other numbers of dimensions, the strategy based on trigonometric approximations for Jacobi polynomials at large order is not immediately viable.

The Bessel-type formulas guarantee uniform convergence near one of the two ends of the interval $[-1,1]$, usually chosen to be $x=1$ in the expositions. These formulas can likewise be retrieved from \cite{DLMF} or the original publications \cite{BG,WZ}. A drawback of resorting to the Bessel function expressions is that they are much less straightforwardly integrated. For that reason, even when the individual polynomials inside the integral in (\ref{eq:C2}) have been accurately approximated, extracting the leading behavior of the integral still presents a challenge. We shall see, however, that this evaluation scheme can work successfully when all the polynomial degrees in (\ref{eq:C2}) are sent to infinity. (This is, however, not the only limit one might be interested to consider in the context of AdS stability studies.) A useful observation to keep in mind is that uniform convergence for individual polynomials near $x=1$ (this corresponds to the origin in the coordinates normally used in the literature for spherically symmetric AdS perturbations) is sufficient to ensure that the approximation can be used inside the integral (\ref{eq:C2}). This is because the other end of the integration range, $x=-1$ (corresponding to the AdS boundary) is suppressed by the factor $(1+x)^{d-1}$, and the uniformity of convergence there turns out to play no significant role.

We finally mention the approach of \cite{int}, quite different in spirit and aiming at direct evaluation of the asymptotics of integrals of the form $\int dx\, \mu(x)\, x^\nu\,p_m(x)p_n(x)$ for any family of orthogonal polynomials $p_m(x)$ with respect to the measure $\mu(x)$. This is in contrast to the previous approaches we have described, giving asymptotic expressions for the polynomials themselves, which, with luck, can be converted to asymptotic expressions for their integrals.

The approach of \cite{int} can be summarized as follows. One first introduces ``wavefunctions'' $\psi_n(x)\equiv (\mu(x)/h_n)^{1/2} p_n(x)$, where the normalization constants $h_n$ are chosen in such a way that $\int dx\,\psi_m\psi_n=\de_{mn}$. From the usual recurrence relation for orthogonal polynomials, one then concludes that
\beq
x\psi_n(x)=a_{n-1}\psi_{n-1}(x)+b_n\psi_n(x)+a_n\psi_{n+1}(x),
\label{recurs}
\eeq
which can be equivalently rewritten as
\beq
x\psi_n(x)= \sum_m J_{nm} \psi_m(x).
\eeq
$J$ is simply a semi-infinite symmetric tridiagonal matrix with $b_n$ on the main diagonal and $a_n$ both immediately above and below the main diagonal  ($J_{nn}\equiv b_n$, $J_{n,n+1}=J_{n+1,n}\equiv a_n$). One can then use this representation to write
\beq
 \int dx \,\mu(x)\, x^\nu p_m(x)p_n(x)=\sqrt{h_mh_n} \int dx \, x^\nu \psi_m(x)\psi_n(x)=\sqrt{h_mh_n}[J^\nu]_{mn},
\label{overlap}
\eeq
where $J^\nu$ is the matrix $J$ raised to the power $\nu$ with ordinary matrix multiplication. Since $a_n$ and $b_n$ are explicitly known for standard families of orthogonal polynomials, including the Jacobi polynomials, their large $n$ asymptotics can be explicitly recovered, whereupon the limit of $[J^\nu]_{mn}$ for large $m$ and $n$ can be constructed, completing the evaluation of (\ref{overlap}). (A disadvantage of these approach is that subleading $1/n$ and $1/m$ corrections are difficult to handle systematically, even though an algorithm for constructing them is in principle described in \cite{int}. These corrections may become important when the simple power $x^\nu$ in (\ref{overlap}) becomes replaced by a high order polynomial with large coefficients of individual powers, as happens in some interaction coefficient expressions.)

In what follows, we shall show how some of the methods presented above can be used in practice to evaluate the ultraviolet asymptotics of (\ref{eq:C2}). We shall see that there are subtleties involved and different methods have to be employed in different situations. Our approach will be ultimately pragmatic: rather than addressing all the complications in a fashion rigorous by the standards of mathematical literature, we shall make, in each case, plausible assumptions on the features determining the asymptotic behavior, and verify our derivations by comparisons with numerical evaluations of the interaction coefficients. We shall see that, when two or three indices in (\ref{eq:C2}) are taken to be large, an approach along the lines of \cite{int} based on an application of (\ref{overlap}) works well. This approach fails, however, for reasons we shall explain below, when all the four indices in (\ref{eq:C2}) become large. In that case (unlike for two or three large indices) using the Bessel function asymptotics for the individual polynomials in (\ref{eq:C2}) works well and recovers the correct ultraviolet behavior.


\section{Interactions involving two ultraviolet modes}

When one considers $C_{klmn}$ of (\ref{eq:C2}) in which only two indices, say, $m$ and $n$ become large, the approach of \cite{int} centered on the application of (\ref{overlap}) is extremely suitable.

Expressions for the recursion coefficients in (\ref{recurs}) for many families of orthogonal polynomials are given in \cite{int}. For the Jacobi polynomials $P^{(\alpha,\beta)}_n$ orthogonal with respect to the measure $(1-x)^\alpha(1+x)^\beta$, these expressions take the form
\begin{align}
&a_n=\frac{\sqrt{(n+1)(n+1+\alpha)(n+1+\beta)(n+1+\alpha+\beta)}}{2(n+1+(\alpha+\beta)/2)\sqrt{(n+(\alpha+\beta+1)/2)(n+(\alpha+\beta+3)/2)}},\\ &b_n=\frac{\beta^2-\alpha^2}{4(n+(\alpha+\beta)/2)(n+1+(\alpha+\beta)/2)}.
\end{align}
These formulas have a very simple large $n$ behavior: $a_n\approx 1/2$, $b_n\approx 0$. According to (\ref{overlap}), the asymptotics of integrals of $x^\nu$ times two large degree Jacobi polynomials can be understood in terms of a simple tridiagonal matrix $J$ with $J_{nn}=0$, $J_{n,n+1}=J_{n+1,n}=1/2$, raised to the power $\nu$. The explicit expression can be written as
\begin{equation} \label{eq:French}
\int_{-1}^{1} dx \, \mu(x)\,x^\nu P^{(\frac{d}{2}-1,\frac{d}{2})}_m(x)P^{(\frac{d}{2}-1,\frac{d}{2})}_n(x)  \simeq 
\begin{cases}
\dsty \frac{2^{d-\nu-1}}{\sqrt{mn}}{\nu \choose \sigma}&\mbox{if}\quad\sigma \equiv\dsty \frac{\nu+n-m}{2}  \in \mathbb{N}\vspace{2mm} \\
\dsty 0 &\mbox{if}\quad \sigma \notin \mathbb{N}
\end{cases}
\end{equation}
for $m>n\gg 1$. We have used $h_n\approx 2^{d-1}/n$, which is valid for Jacobi polynomials at large $n$, as explained in \cite{int}. Note that the integral on the left hand side vanishes by the Jacobi polynomial orthogonality if $m-n>\nu$. It is interesting that the asymptotics is insensitive to the value of $d$ (more generally, with respect to the measure parameters $\alpha$ and $\beta$), except for the overall normalization factor.

In order to analyze the behavior of (\ref{eq:C2}) with two large indices, one simply has to compute the integral $\int dx\,\mu(x) Q(x) P_m(x)P_n(x)$, where $Q(x)=(1+x)^{d-1} P_k(x) P_l(x)$ is a fixed degree polynomial (and we have omitted the measure parameter superscripts on the Jacobi polynomials $P_n$). All one needs is to expand $Q(x)$ explicitly in powers of $x$ and apply (\ref{eq:French}) individually to each power. While this algorithm is entirely sufficient, say, for numerical purposes, for analytic work it is convenient to derive a closed form expression that implements the substitution of (\ref{eq:French}) for individual powers of $x$ in $Q(x)$, for example, via a complex contour integral. This is what we shall turn to next.

First, consider (\ref{eq:French}) for the case when $m-n$ is even, i.e., $m-n=2p$ with $p \in \mathbb{N}$. In this case, only even powers $\nu$ in $Q(x)$, i.e., $\nu=2q$ for $q\in \mathbb{N}$, contribute to the integral. One can explicitly isolate these powers from the onset as $(Q(x)+Q(-x))/2$. We then recall that $(1+z)^{2q} = \sum_{r=0}^{2q} {2q \choose r}z^r$, and construct the following contour integral along an arbitrary closed curve $\gamma$ enclosing the origin in the complex plane
 \begin{equation}
 \frac{1}{2\pi i}\oint_\gamma \left(\frac{1+z}{2\sqrt{z}}\right)^{2q} z^{p-1}dz = 2^{-2q}\,{2q \choose q-p}.
 \end{equation}
With these observations in mind, we can write, for the case of even $m-n$,
\begin{equation} 
 \int_{-1}^{1} \mu(x)Q(x)P_m(x)P_n(x) dx = \frac{2^{d-2}}{\sqrt{mn}}  \oint_{\gamma}  \frac{z^{\frac{m-n}{2}-1}}{2 \pi i} \bigg(Q \left(\frac{1+z}{2\sqrt{z}}\right) + Q \left(-\frac{1+z}{2\sqrt{z}}\right)\bigg) dz.
\label{Qeven}
 \end{equation} 
A similar derivation can be performed for odd $m-n$, with the results combining to a general formula valid for any large $m$ and $n$:
\begin{equation}
 \int_{-1}^{1} \mu(x)Q(x)P_m(x)P_n(x) dx = \frac{2^{d-2}}{\sqrt{mn}} \oint_{\gamma}  \frac{z^{\frac{m-n}{2}-1}}{2 \pi i} \bigg(Q \left(\frac{1+z}{2\sqrt{z}}\right) + (-1)^{m-n} Q \left(-\frac{1+z}{2\sqrt{z}}\right)\bigg) dz.
\label{Qall}
 \end{equation}

The above representation can be used to derive simple, even if a bit cumbersome, expressions in terms of triple finite sums for general resonant interaction coefficients with two small and two large indices, namely, $C_{k,l,m,m+k-l}$ with $m\to\infty$. One starts with (\ref{eq:C2}) and substitutes (\ref{Jacobigen}) for $P_k$ and $P_l$, applying afterwards (\ref{Qall}) and evaluating the resulting simple complex contour integrals explicitly (we give some more details in the next section, where we deal with a similar, but somewhat less obvious setup).
For odd $d$, the result is
\begin{align}
&C_{k,l,m,m+k-l} \simeq  \frac{\sqrt{k!l!(k+d-1)!(l+d-1)!}}{2^{d+2k+2l-3}\pi}(2k+d)!!(2l+d)!! \label{C2odd}\\
& \times \sum_{s=0}^k\sum_{t=0}^l \frac{\sum\limits_{u=0}^{\min(k,l)+\frac{d-1}{2}}{2d+2s+2t-2 \choose 2u}{2k + 2l - 2s - 2t \choose 2\min(k,l)+d-2u-1}-\sum\limits_{u=0}^{\min(k,l)+\frac{d-3}{2}}{2d+2s+2t-2 \choose 2u+1}{2k + 2l - 2s - 2t \choose 2\min(k,l)+d-2u-2} }{s!t!(k-s)!(l-t)!(2k-2s+d-2)!!(2l-2t+d-2)!!(2s+d)!!(2t+d)!!}.\nonumber 
\end{align}
For even $d$, the result is
\begin{align}
&C_{k,l,m,m+k-l} \simeq  \frac{\sqrt{k!l!(k+d-1)!(l+d-1)!}}{2^{2d+2k+2l-4}}(k+d/2)!(l+d/2)! \label{C2even} \\
& \hspace{3mm}\times \sum_{s=0}^k\sum_{t=0}^l \sum_{u=0}^{\min(k,l)+\frac{d}{2}-1}\frac{{2d+2s+2t-2 \choose 2u+1}{2k + 2l - 2s - 2t \choose 2\min(k,l)+d-2u-2}  -{2d+2s+2t-2 \choose 2u}{2k + 2l - 2s - 2t \choose 2\min(k,l)+d-2u-1}}{s!t!(k-s)!(l-t)!(k+\frac{d}{2}-s-1)!(l+\frac{d}{2}-t-1)!(s+\frac{d}{2})!(t+\frac{d}{2})!}.\nonumber
\end{align} 
In the above formulas, $\min(k,l)$ denotes the smaller of $k$ and $l$, and $!!$ is the usual `semifactorial' understood as the product of all odd numbers less than or equal to its (odd) argument.

We have checked the above asymptotic expressions against a few actual numerically computed values, including $C_{5,13,700,708}$ in 3 dimensions, $C_{8,17,800,809}$ in 4 dimensions, $C_{2,20,850,868}$ in 5 dimensions and $C_{9,18,900,909}$ in 6 dimensions. We obtained the values $16.5129$, $286.591$, $253.568$, and $27291.8$, respectively, while our formulas predict the asymptotic values $16.5129$, $286.610$, $253.588$, and $27297.9$, respectively.   


\section{Interactions involving three ultraviolet modes}\label{3ind}

Interaction coefficients with three large indices play a key role \cite{CEJV} in determining properties of quasiperiodic solutions to the time-averaged effective equations for AdS perturbations. These solutions, in which the initial conditions are fine-tuned in such a way that the energy does not flow between different modes at all, have spectra strongly localized around one normal mode. Interactions of this mode with three ultraviolet modes dominate the equations determining the shape of quasiperiodic spectra. In this section, we shall show how the techniques displayed above for interactions involving two ultraviolet modes allow one to derive closed form expressions for the three ultraviolet mode asymptotics.

To keep the formulas more compact, we shall start with the interaction coefficient $C_{0,j,k,k-j}$ relevant for quasiperiodic solutions centered on mode 0, and look for an asymptotic expression when the  values of $j$, $k$ and $k-j$ are all large. At the end of this section, we give the asymptotic results with mode 0 replaced by any other fixed mode.

The relevant interaction coefficient can be written as
\begin{equation} \label{eq:3large}
C_{0,j,k,k-j} = \frac{k_0k_jk_kk_{k-j}}{2^{2d}}\int_{-1}^{1} \mu(x)Q(x)P^{(\frac{d}{2}-1,\frac{d}{2})}_j(x)P^{(\frac{d}{2}-1,\frac{d}{2})}_k(x)dx, 
\end{equation}
where $Q(x) = (1+x)^{d-1}P^{(\frac{d}{2}-1,\frac{d}{2})}_{k-j}(x)$. In order to evaluate this integral asymptotically, we can resort to the expression (\ref{Qall}) in the complex plane. In contrast to the derivations of the previous section, $Q(x)$ is now itself a high degree polynomial, for which one needs to employ some approximation in order to handle it efficiently. We find the following complex plane representation, due to \cite{Elliott}, useful:
\beq
P^{(\frac{d}{2}-1,\frac{d}{2})}_n(z) \simeq \frac{\Gamma(2n+d)}{2^{2n+d/2}\Gamma(n+1)\Gamma(n+d)}\frac{(z+\sqrt{z^2-1})^{n+d/2}}{(z-1)^{(d-1)/4}(z+1)^{(d+1)/4}}.
\label{PElliott}
\eeq
This formula is uniformly valid on any curve in the  $z$-plane lying outside the real interval $[-1,1]$. Substituting this formula into the polynomial $Q(x)$ in (\ref{eq:3large}), applying (\ref{Qall}), and utilizing Stirling's approximation for all $\Gamma$-functions of large arguments, we arrive at an integral of the following form: 
\begin{equation}
C_{0,j,k,k-j} \simeq  \frac{1}{2^{d-2}\Gamma(\frac{d}{2})} \sqrt{\frac{(d-1)!}{\pi}}\frac{1}{2\pi i} \oint_{\gamma} \bigg( \frac{(1+\sqrt{z})^{2d-3}+(-1)^{d-1}(1-\sqrt{z})^{2d-3}}{(1-z)^{\frac{d-1}{2}}z^{\frac{d+1}{2}}} \bigg) dz.
\end{equation}
(Note that for a contour $\gamma$ close to the origin in the $z$-plane in (\ref{Qall}), the argument of $Q$, and hence of the Jacobi polynomial contained in $Q$, has a very large absolute value, justifying the use of the above approximation for Jacobi polynomials.) For odd $d$, the integrand can be expanded as 
 \beq
  \frac{(1+\sqrt{z})^{2d-3}+(-1)^{d-1}(1-\sqrt{z})^{2d-3}}{(1-z)^{\frac{d-1}{2}}z^{\frac{d+1}{2}}} = 2\sum_{r=0}^{d-2}\sum_{s=0}^{\infty} {2d-3 \choose 2r} {\frac{d-3}{2}+s \choose s} \frac{1}{z^{\frac{d+1}{2}-r-s}}.
 \eeq
 The complex integral will simply pick the coefficient in front of $1/z$, i.e., $\frac{d+1}{2}-r-s =1$. We finally arrive at the following asymptotic expression, valid for odd $d$:
\begin{equation}
C_{0,j,k,k-j} \simeq \frac{1}{2^{d-3}\Gamma(\frac{d}{2})} \sqrt{\frac{(d-1)!}{\pi}}\,\,\displaystyle {\sum_{r=0}^{\frac{d-1}2}} {2d-3 \choose 2r} {d-r-2 \choose \frac{d-1}{2}-r}.
\end{equation} 
For even $d$, one can use the same procedure to obtain the corresponding asymptotic expression:
\begin{equation}
C_{0,j,k,k-j} \simeq \frac{1}{2^{d-3}\Gamma(\frac{d}{2})} \sqrt{\frac{(d-1)!}{\pi}}\,\,\displaystyle {\sum_{r=0}^{\frac{d}2-1}} {2d-3 \choose 2r+1} {d-r-\frac{5}{2} \choose \frac{d}{2}-r-1}.
\end{equation} 
The `binomial coefficients' above are understood to be given by (\ref{binom}) and do not require integer values of the arguments. Using $\Gamma(n+1/2)=\sqrt{\pi}(2n)!/4^nn!=\sqrt{\pi}(2n-1)!!/2^n$, one can further simplify the above formulas to
\beq
C_{0,j,k,k-j} \simeq \frac{4\left((d-1)/2\right)!}{\pi\sqrt{(d-1)!}}\,\,\displaystyle {\sum_{r=0}^{\frac{d-1}2}} {2d-3 \choose 2r} {d-r-2 \choose \frac{d-1}{2}-r}
\label{Codd}
\eeq
for odd $d$, and
\newpage
\beq
C_{0,j,k,k-j} \simeq \frac{8(d-1)}{ \sqrt{\pi(d-1)!}}\,\,\displaystyle {\sum_{r=0}^{\frac{d}2-1}} {2d-3 \choose 2r+1} \frac{(2d-2r-5)!!}{2^{d-r}(d/2-r-1)!}
\label{Ceven}
\eeq
for even $d$. The semifactorial, which we have already encountered in (\ref{C2odd}), and which is denoted by $!!$, appears once again in the above formula.

We have verified the convergence of $C_{0,j,k,k-j}$  to these asymptotic values numerically at $d=3,4,5,6$, as can be seen, for instance, from the plots given below:\vspace{5mm}

\noindent For $d=3$: $C_{0,j,k,k-j} \simeq \frac{8\sqrt{2}}{\pi} $
\begin{center}
\includegraphics[scale=0.5]{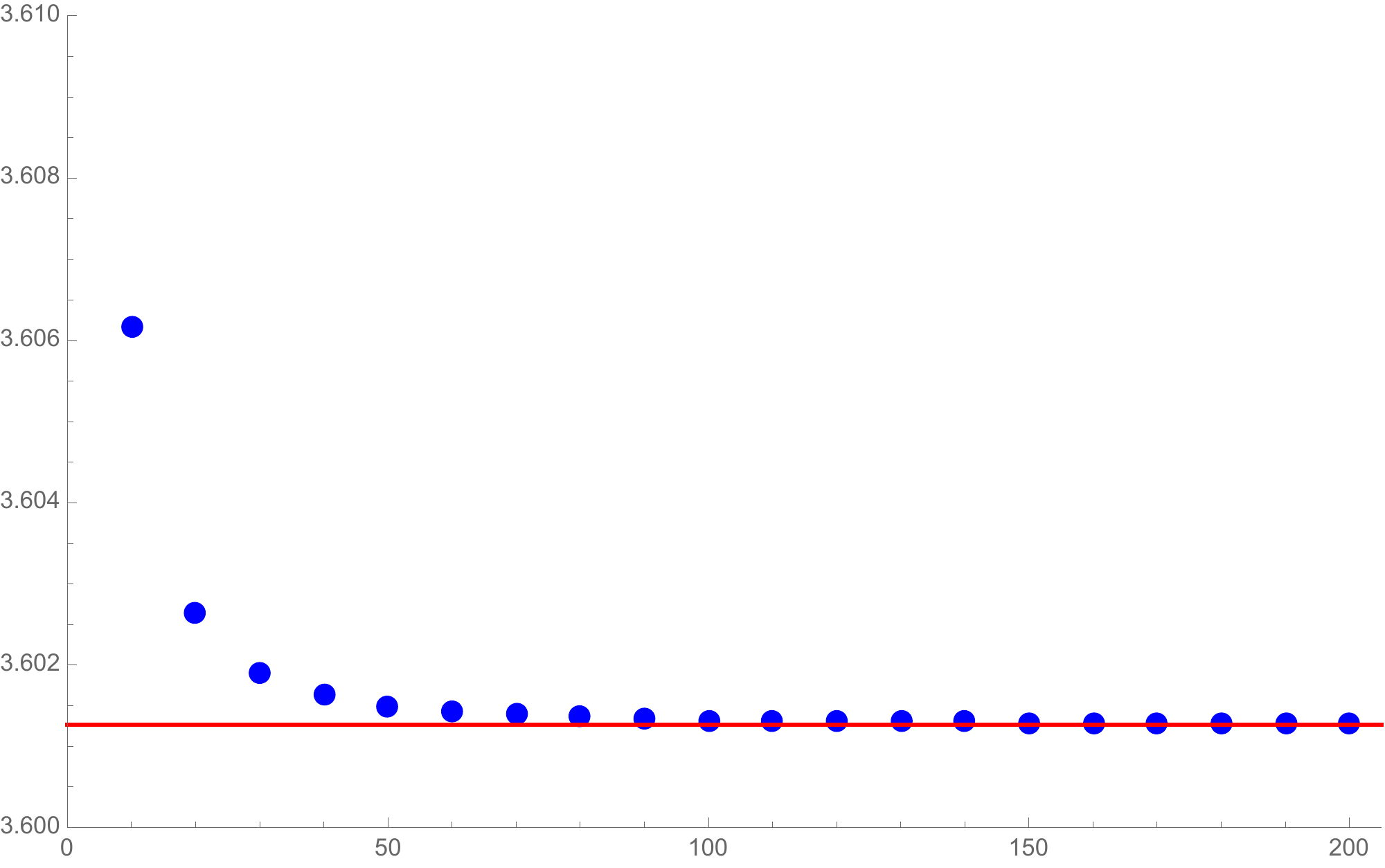} \\
(Blue) $C_{0,a,2a,3a}$ plotted against $a$, (Red) the asymptotic value $\frac{8\sqrt{2}}{\pi}$.
\end{center}

\noindent For $d=4$: $C_{0,j,k,k-j} \simeq \frac{35}{2}\sqrt{\frac{3}{2\pi}} $
\begin{center}
\includegraphics[scale=0.5]{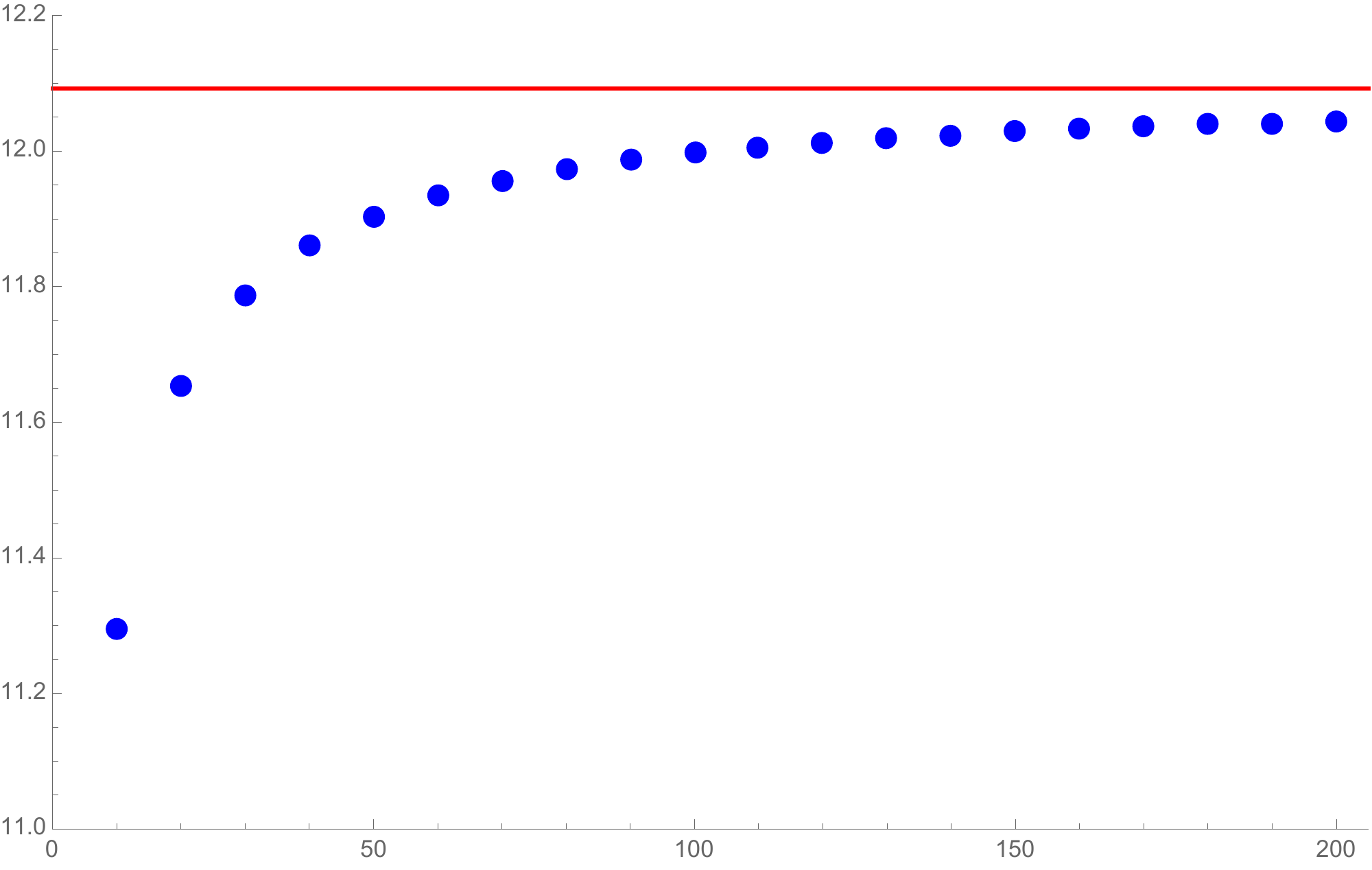} \\
(Blue) $C_{0,a,2a,3a}$ plotted against $a$, (Red) the asymptotic value $\frac{35}{2}\sqrt{\frac{3}{2\pi}}$.
\end{center}
\newpage
\noindent For $d=5$: $C_{0,j,k,k-j} \simeq \frac{160}{\pi}\sqrt{\frac{2}{3}} $
\begin{center}
\includegraphics[scale=0.5]{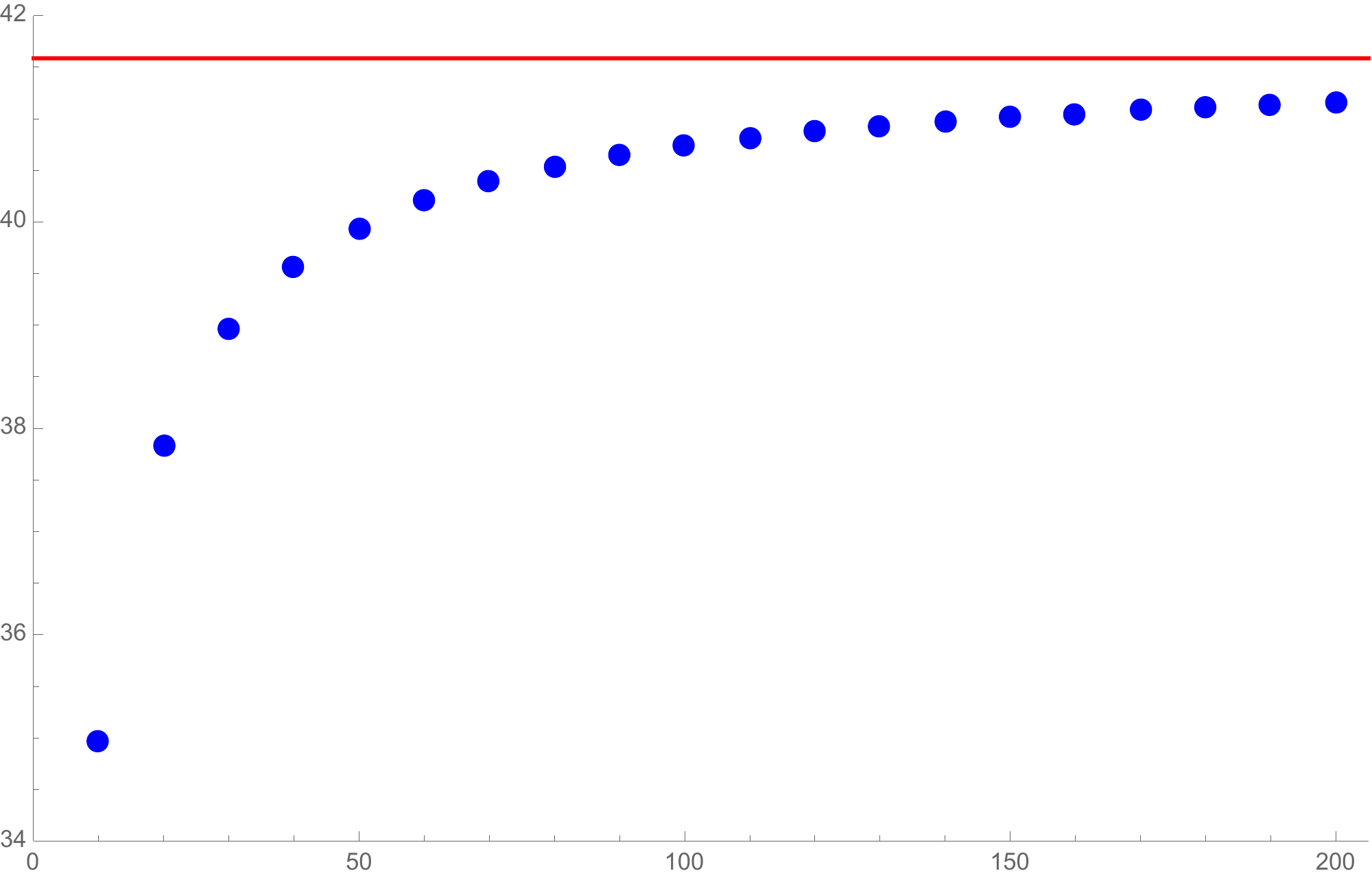} \\
(Blue) $C_{0,a,2a,3a}$ plotted against $a$, (Red) the asymptotic value $\frac{160}{\pi}\sqrt{\frac{2}{3}}$.
\end{center}

\noindent For $d=6$: $C_{0,j,k,k-j} \simeq \frac{3003}{32}\sqrt{\frac{15}{2\pi}}$
\begin{center}
\includegraphics[scale=0.5]{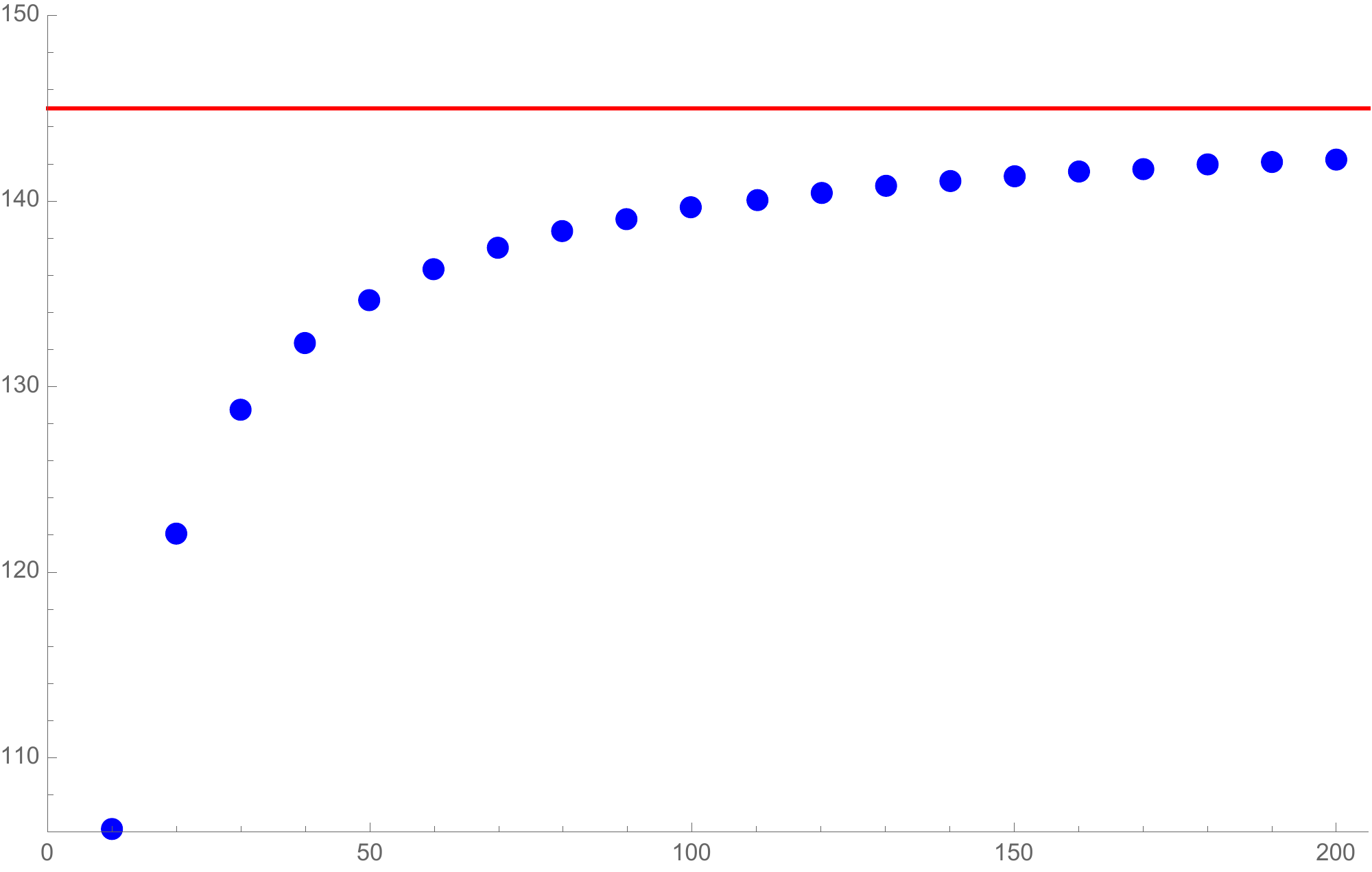} \\
(Blue) $C_{0,a,2a,3a}$ plotted against $a$, (Red)the asymptotic value $\frac{3003}{32}\sqrt{\frac{15}{2\pi}}$.
\end{center}\vspace{5mm}

It is possible to repeat the above procedure for a general resonant coefficient $C_{n,j,k,k-j+n}$ when $n$ is fixed and all of the other indices become large. This is done by substituting (\ref{Jacobigen}) for $P_n$, which results in a sum over $s$ varying between $0$ and $n$.
One can then apply (\ref{Qall}) to each term in the sum individually using the following specification:
\begin{equation}
Q(x) = (1+x)^{d-1+s}(x-1)^{n-s}P^{(\frac{d}{2}-1,\frac{d}{2})}_{k-j+n}(x).
\end{equation}
Afterwards, a slightly more involved version of the procedure we have employed above at $n=0$ can be used to extract the general asymptotics.

For odd $d$, one obtains
\begin{align}
C_{n,j,k,k-j+n}\simeq&\frac{\sqrt{n!(n+d-1)!}}{2^{2n+\frac{d-5}{2}}\pi(2n+d-2)!!}\hspace{-2mm}\sum_{q=0}^{\mbox{\footnotesize min}\left(\frac{d-3}{4},n\right)}\sum_{r=0}^{\frac{d-1}{2}+n} \frac{(2n+d)!!(2n+d-2)!!}{q!(n-q)!(2n-2q+d)!!(2q+d-2)!!} \nonumber \\
&\hspace{2cm} \times {2d-3+2n-4q \choose 2r}{d-2+n-2q-r \choose \frac{d-1}{2}+n-r}\label{C3odd}
\end{align} 
In the notation of (\ref{Jacobigen}), $q$ corresponds to $n-s$ , and the summation over $q$ runs up to the largest integer not greater than either $(d-3)/4$ or $n$.
For even $d$, one obtains
\begin{align}
C_{n,j,k,k-j+n}\simeq&\frac{\sqrt{n!(n+d-1)!}}{2^{3n+\frac{3d}{2}-4}\sqrt{\pi}(n+\frac{d}{2}-1)!}\sum_{s=0}^{n}\sum_{r=0}^{\frac{d}{2}+n-1}   \frac{2^r(2d+4s-2n-2r-5)!!}{(\frac{d}{2}+n-r-1)!(d+4s-4n-3)!!} \nonumber \\
&\hspace{2cm} \times{n+\frac{d}{2}-1 \choose s}{n+\frac{d}{2} \choose n-s} {2d-3+4s-2n \choose 2r+1}.\label{C3even}
\end{align} 
When the third binomial coefficient in the above formula appears with a negative value of the upper argument, it is understood as
\begin{equation}
{-m \choose r} \equiv (-1)^r{m+r-1\choose r}.
\end{equation} 

 We have checked the above asymptotics numerically for a few randomly chosen values of the indices in $C_{n,j,k,k-j+n}$ in a few different dimensions, including $C_{4,700,500,204}$ in 3 dimensions,
$C_{3,850,520,233}$ in 4 dimensions, $C_{2,900,600,302}$ in 5 dimensions, and $C_{5,900,400,505}$ in 6 dimensions. We have obtained the actual numerical values  $13.9477$, $84.37$, $286.714$ and $9796.27$, respectively, while the above formulas predict the asymptotic values $13.9476$, $84.85$, $289.896$ and $10049.8$, respectively. 

A somewhat surprising feature of our asymptotic results is that the asymptotic value of $C_{n,j,k,k-j+n}$ is completely independent of $j$ and $k$, provided that $j$, $k$ and $k-j+n$ are all large. This flat `plateau topography' of the plot of $C_{n,j,k,k-j+n}$ as a function of $j$ and $k$ is familiar from the analysis of \cite{CEJV}, which was only applicable to $d=3$. Our current considerations display it in all dimensions.


\section{Interactions involving only ultraviolet modes}\label{4ind}

For the interaction coefficients with four large indices, more specifically, the limit of $a \rightarrow \infty$ in $C_{aj,ak,al,am}$, the methods we have described above cannot be employed efficiently. One could have tried to repeat the derivation under (\ref{eq:3large}) with $Q(x)$ containing two large degree Jacobi polynomials instead of one. One discovers however, that the resulting asymptotic expressions do not stand numerical tests. Investigations into the causes of the failure reveal that, in particular, subleading corrections in (\ref{PElliott}), which are known explicitly, become important. (This happens because, after integration, corrections to individual polynomials proportional to, say, $1/n$ get multiplied with another index, producing a contribution that survives under uniform scaling of the indices to infinity.) By symmetry of the interaction coefficients under index permutation, this would imply that subleading corrections to (\ref{eq:French}), for which there are no simple closed expressions, also become important, which effectively drives evaluation attempts of this sort to a dead end.

We have experimented with a number of evaluation strategies based on different combinations of asymptotics along the lines of \cite{Elliott,Hahn} and \cite{int}, before realizing that the following simple-minded approach works well specifically when all the indices in the interaction coefficients are taken to be large. We first recall the following asymptotic expression for the Jacobi polynomials \cite{DLMF, BG,WZ}:
\begin{equation}
P^{(\frac{d}{2}-1,\frac{d}{2})}_n(\cos 2\xi) \simeq  \frac{\sqrt{\xi}}{(\sin \xi)^{(d-1)/2}(\cos \xi)^{(d+1)/2}}\,J_{\frac{d}{2}-1}(\omega_{n}\xi).
\end{equation} 
where $\omega_n = 2n+d$, and $J_{\frac{d}{2}-1}(\omega_{n}\xi)$ are Bessel functions. One important property of this asymptotic formula is that the large $n$ convergence is uniform near $\xi=0$ (the center of spherical symmetry of the AdS perturbations we are considering). The interaction coefficients with four large indices can thus be written as
\begin{equation}
C_{aj,ak,al,am} \simeq k_{aj}k_{ak}k_{al}k_{am}\int_{0}^{\frac{\pi}{2}} \xi^2 \,\frac{\cos^{d-3}\xi}{\sin^{d-1}\xi} J_{\frac{d}{2}-1}(\omega_{aj}\xi)J_{\frac{d}{2}-1}(\omega_{ak}\xi)J_{\frac{d}{2}-1}(\omega_{al}\xi)J_{\frac{d}{2}-1}(\omega_{am}\xi) d\xi.
\label{Besselint}
\end{equation}
The above integral expression is still rather unwieldy and does not immediately tell what the large $a$ asymptotics is, until one makes the following key observation. The Bessel functions in the above integral oscillate very rapidly in the entire range as $a$ increases. At the same time, it is known from numerical evaluations and from the results of \cite{CEJV} at $d=3$ that the interaction coefficients grow with $a$. The only way such growing behavior can emerge is due to the rapid variation of $1/\sin^{d-1}\xi$ near $\xi=0$. Any region where $1/\sin^{d-1}\xi$ behaves regularly can only produce decaying contributions at large $a$ in (\ref{Besselint}). It is thus natural to assume (and this assumption can be further verified numerically, for instance) that ${\cos^{d-3}\xi}/{\sin^{d-1}\xi}$ in (\ref{Besselint}) can be replaced by $1/\xi^{d-1}$ while the upper limit of integration can be pushed to infinity, without affecting the large $a$ asymptotics. 
The resulting expression reads
\begin{equation}
C_{aj,ak,al,am} \simeq k_{aj}k_{ak}k_{al}k_{am}\int_{0}^{\infty} \frac{d\xi}{\xi^{d-3}}\,\,J_{\frac{d}{2}-1}(\omega_{aj}\xi)J_{\frac{d}{2}-1}(\omega_{ak}\xi)J_{\frac{d}{2}-1}(\omega_{al}\xi)J_{\frac{d}{2}-1}(\omega_{am}\xi).
\end{equation}
At this stage, $a$ can be simply scaled out from the integral, leaving a manifest asymptotic formula at $a \rightarrow \infty$ (note that, according to (\ref{klarge}), each of the $k$-factors contributes $\sqrt{a}$):
\begin{equation} \label{eq:4large}
\frac{C_{aj,ak,al,am}}{a^{d-2}} \simeq 2^d \sqrt{jklm}  \int_{0}^{\infty} \frac{d\xi}{\xi^{d-3}}\,\, J_{\frac{d}{2}-1}(j\xi)J_{\frac{d}{2}-1}(k\xi)J_{\frac{d}{2}-1}(l\xi)J_{\frac{d}{2}-1}(m\xi).
\end{equation}
An alternative closed form expression for this integral valid for odd $d$ can be found in the appendix.

We would like to point out that the techniques we have presented in this section will work when all the mode number indices in the interaction coefficients are simultaneously taken to infinity, but we see no immediate way to apply them as an alternative approach to the derivations of the previous sections. Indeed, say, for the case of three large indices, one could write an analog of (\ref{Besselint}), but we see no reason why, for that case, the integral should be dominated by one end point of the integration region, and consequently see no obvious way to isolate the leading $a$-dependence (which should approach a constant at large $a$ for that case). Equally non-obvious is how one could convert the Bessel function integrals to the closed form expressions (\ref{C3odd}-\ref{C3even}).

Our aim in the present derivations has been to focus on the infinite $a$ limit and display the relevant asymptotic formulas. Discussing the rate of convergence to these formulas would take us far afield and probably require qualitatively different methods. From our numerical verifications of (\ref{eq:4large}), it appears that convergence becomes slower at higher values of $d$, in a way suggestive of corrections of the form $d/a$, etc. We display below some of our plots for the behavior of $C_{aj,ak,al,am}/a^{d-2}$ at $d=3,4,5,6$.\vspace{7mm}

\noindent For $d=3$: $\frac{C_{a,2a,3a,4a}}{a} \simeq 2.54648 $
\begin{center}
\includegraphics[scale=0.5]{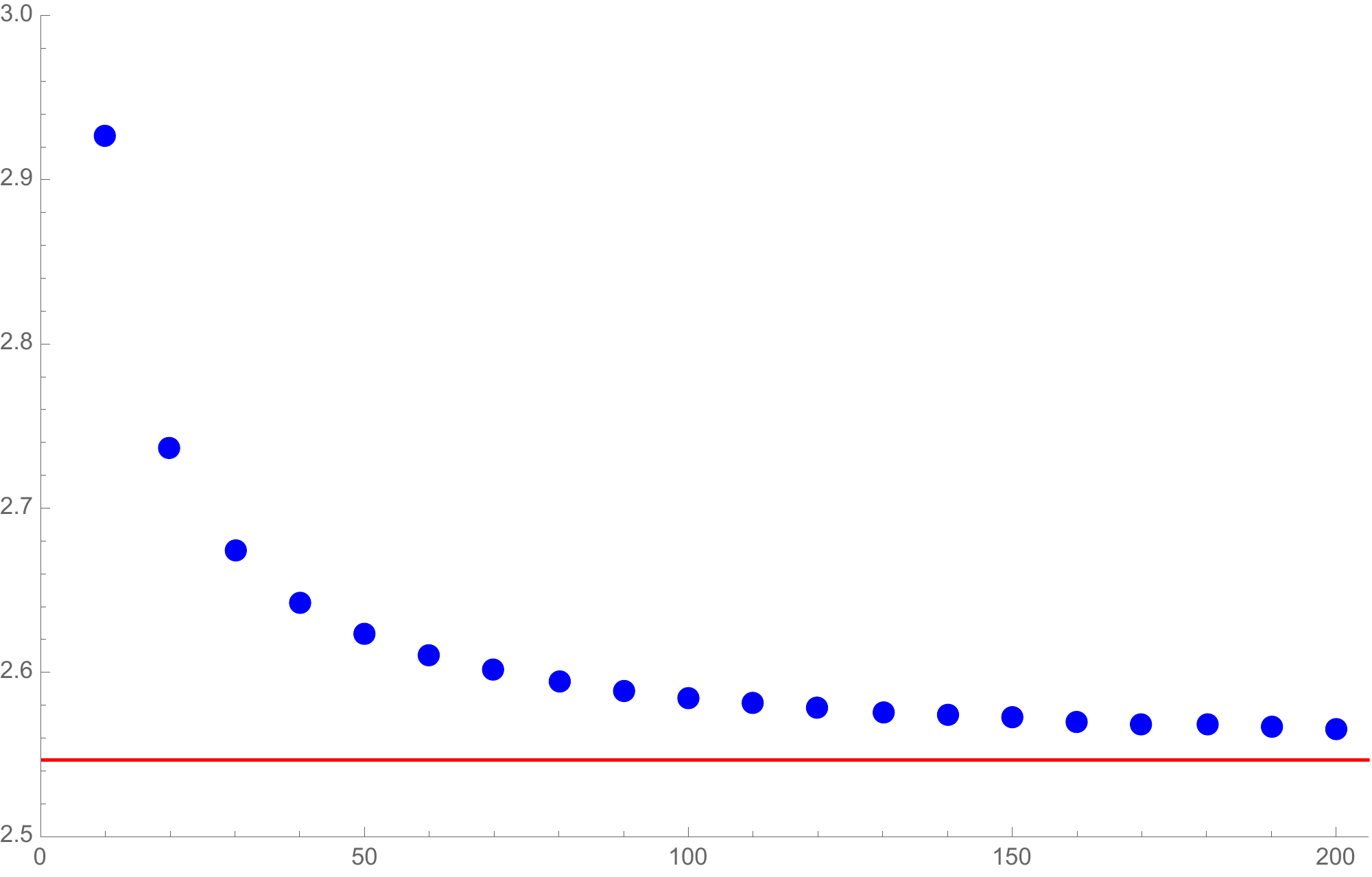} \\
(Blue) $\frac{C_{a,2a,3a,4a}}{a}$ plotted against $a$, (Red) the asymptotic value $2.54648$.
\end{center}

\noindent For $d=4$: $\frac{C_{a,2a,3a,4a}}{a^2} \simeq 2.79640 $
\begin{center}
\includegraphics[scale=0.5]{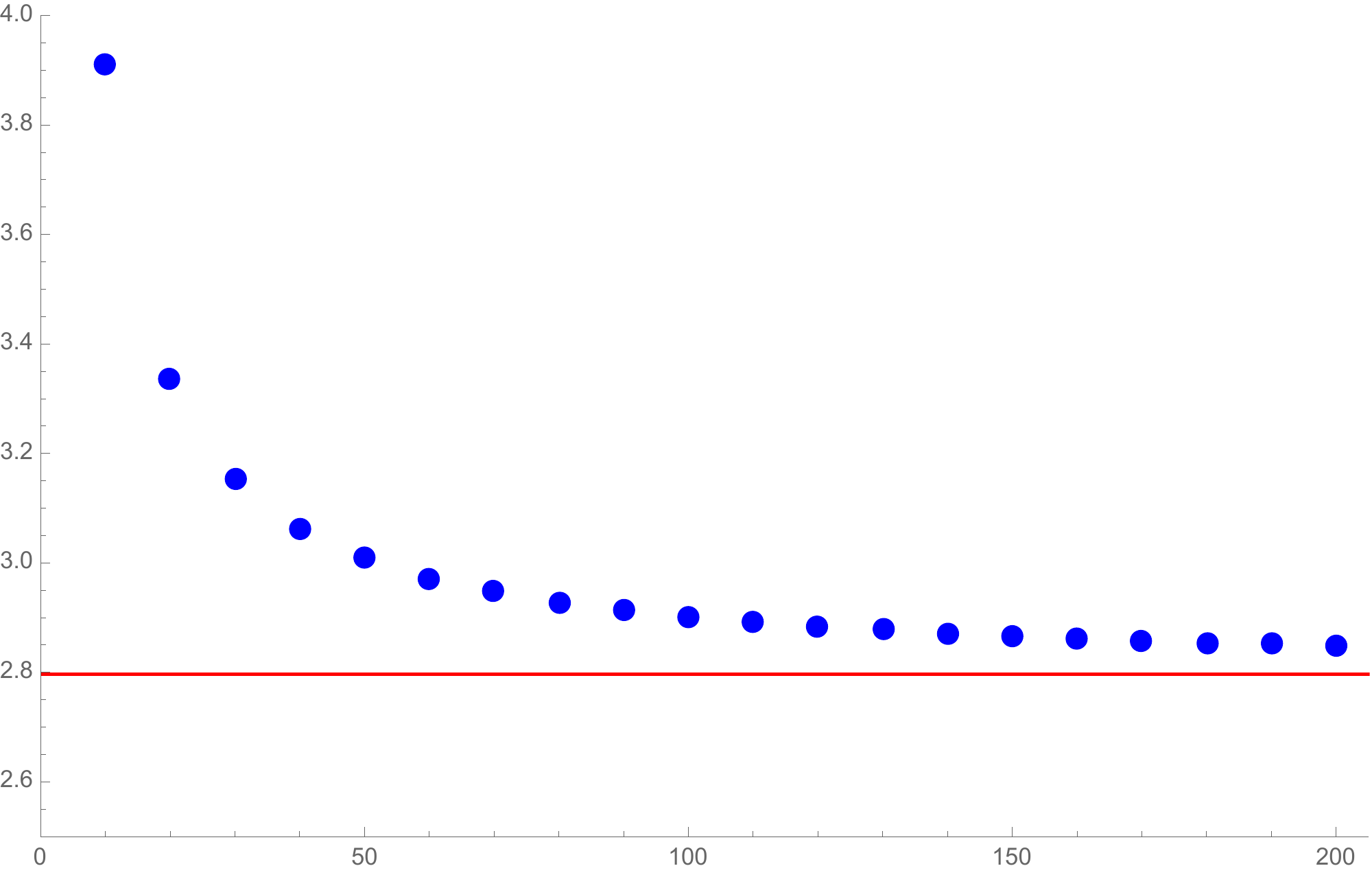} \\
(Blue) $\frac{C_{a,2a,3a,4a}}{a^2}$ plotted against $a$, (Red) the asymptotic value $2.79640$.
\end{center}
\newpage
\noindent For $d=5$: $\frac{C_{a,2a,3a,4a}}{a^3} \simeq 2.18269 $
\begin{center}
\includegraphics[scale=0.5]{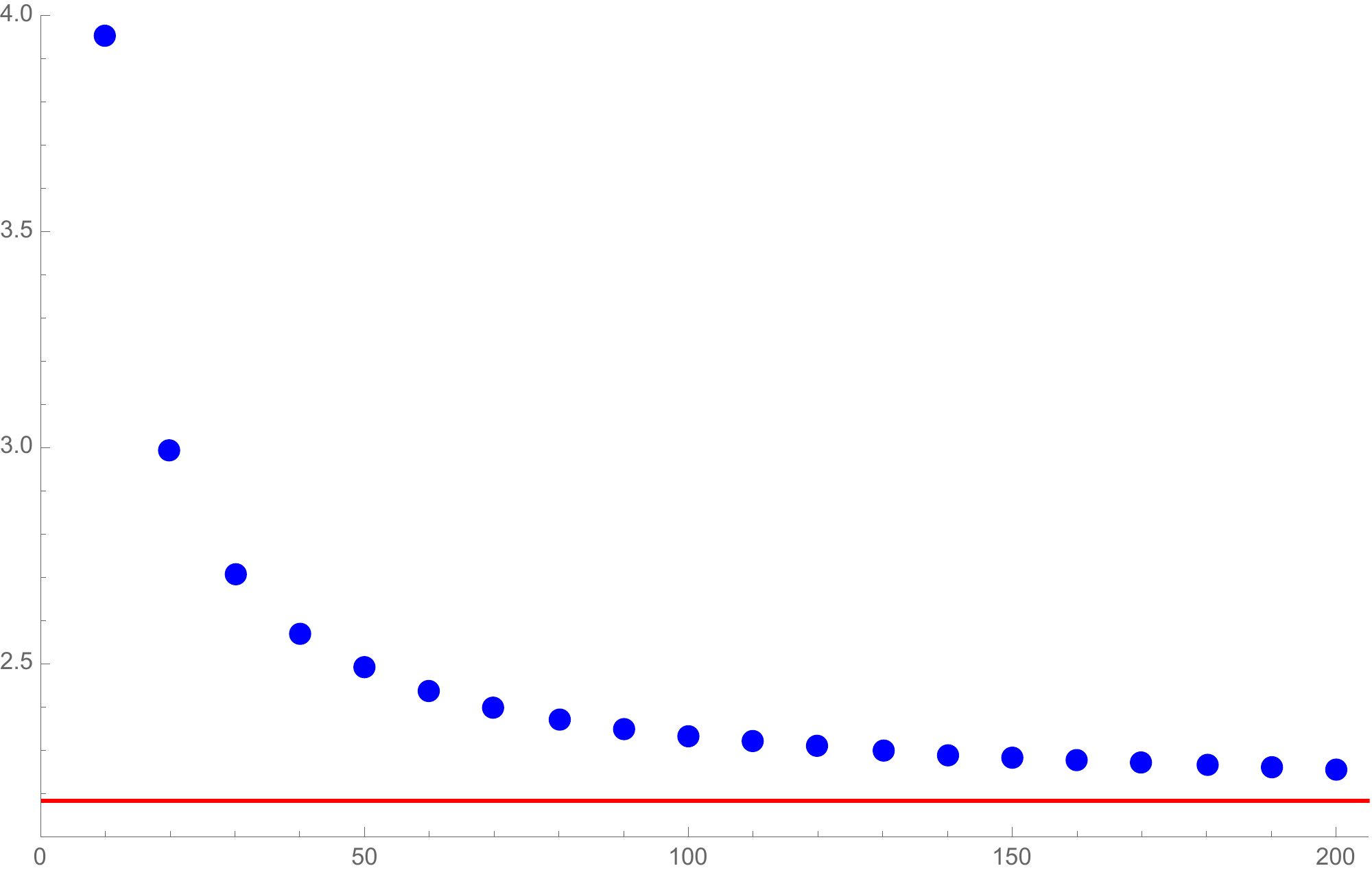} \\
(Blue) $\frac{C_{a,2a,3a,4a}}{a^3}$ plotted against $a$, (Red) the asymptotic value $2.18269$.
\end{center}

\noindent For $d=6$: $\frac{C_{a,2a,3a,4a}}{a^4} \simeq 1.29981 $
\begin{center}
\includegraphics[scale=0.5]{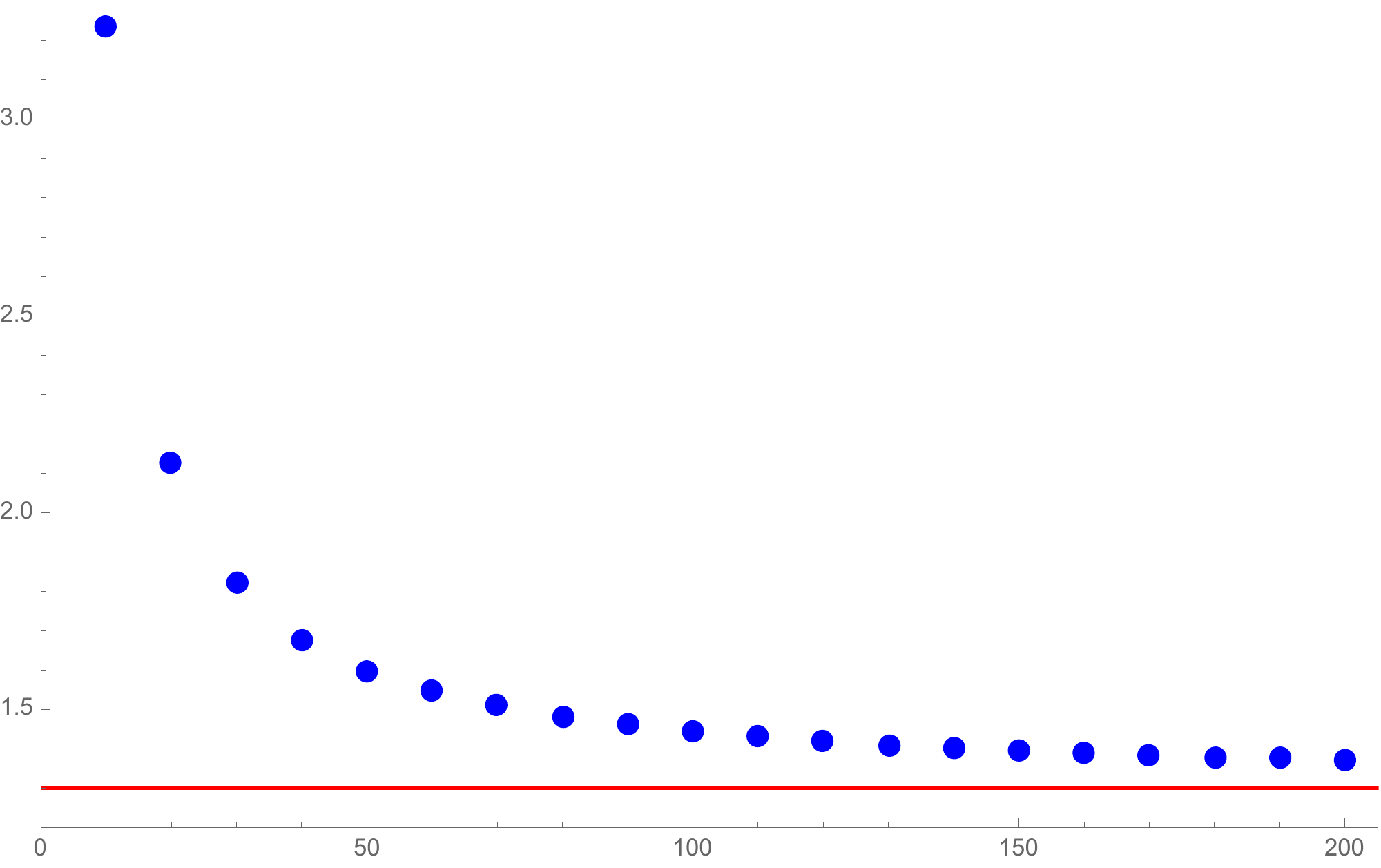} \\
(Blue) $\frac{C_{a,2a,3a,4a}}{a^4}$ plotted against $a$, (Red) the asymptotic value $1.29981$.
\end{center}


\section{Implications for AdS perturbations}

Our main objective has been to develop methodology for asymptotic problems arising in the analysis of AdS perturbations. The previous sections have described general strategies that can be used, and applied them to the simplest case, taken to be a spherically symmetric massless self-interacting probe scalar field in AdS$_{d+1}$. Detailed asymptotic formulas resulted from these considerations. We are not aiming at any in-depth explorations of the dynamical consequences of these asymptotic formulas, and leave this subject to future work. We shall summarize here, however, the general role of the particular asymptotic structures we have explored for studies of small AdS perturbations.

As we have already remarked, the asymptotics for the interaction coefficients (\ref{eq:C1}) with three large and one small indices (considered in section \ref{3ind}) determines the structure of special quasiperiodic AdS perturbations, in which there is no significant energy transfer between normal modes, despite the presence of nonlinearities. In \cite{CEJV}, this problem was analyzed at $d=3$ (which is possible using very basic methods for the asymptotic analysis involved, unlike the cases we have confronted in this paper). It was seen that the interaction coefficients approach a universal limit when three indices are taken large, and this limit only depends on the remaining small index. This flat plateau-like topography of the dependence of the interaction coefficients on mode numbers allowed a fairly complete analysis of quasiperiodic solutions for a nonlinear probe scalar field in \cite{CEJV}. Our current investigations established a similar plateau topography in higher dimensions, meaning that the analytic considerations of \cite{CEJV} will translate to higher dimensions essentially unaltered.

The interaction coefficients (\ref{eq:C1}) with four large indices (considered in section \ref{4ind}) play a central role in a rather different dynamical regime of AdS perturbations. Numerical simulations of \cite{BMR} indicate that solutions to time-averaged equations describing the dynamics of gravitational AdS perturbations develop a singularity in finite time. At this singular moment, a power law perturbation spectum emerges (even if one had started with a spectrum strongly suppressed in the ultraviolet), which is a concrete realization of a turbulent cascade effectively transferring energy to modes of arbitrarily short wavelength. Couplings of high frequency modes among themselves dominate the equations in the regime when the spectrum develops a power-law ultraviolet tail \cite{BMR,CEV3}. The precise structure of the equations is still waiting to be fully explored in this regime, and our asymptotic methods (in application to gravitational nonlinearities) will provide a welcome input.

We note additionally that evaluation of large mode number interaction coefficients is known to constitute the main computational burden in numerical simulations based on time-averaged AdS perturbation theory of the sort undertaken in \cite{BMR,GMLL}. Having accurate formulas controlling the large mode number asymptotics can be used to drastically simplify precisely this resource-demanding aspect of numerical simulations.


\section{Acknowledgments}

We would like to thank Ben Craps, Chethan Krishnan, Rongvoram Nivesvivat and Joris Vanhoof for collaboration on related subjects, and the JHEP anonymous referee for providing a number of stimulating suggestions.
The work of O.E. is funded under CUniverse
research promotion project by Chulalongkorn University (grant reference CUAASC). P.J. is supported by a scholarship under the Development and Promotion of Science and
Technology Talents Project (DPST) by the government of Thailand.


\appendix

\section{Closed form asymptotic expressions for odd $d$}

We quote here, without derivation, an alternative closed form expression for the asymptotic formula (\ref{eq:4large}) involving four ultraviolet modes, valid at odd $d$:
\begin{align}
 \frac{C_{aj,ak,al,am}}{a^{d-2}} & \simeq  \frac{2^{d-2}}{\pi}\bigg[ (-1)^{\frac{d-1}{2}}\chi_{\sssty 1,1,1,1}(j,k,l,m)  +(-1)^{\frac{d-1}{2}}\chi_{\sssty 1,-1,-1,1}(j,k,l,m)\nonumber \\
&+(-1)^{\frac{d-1}{2}}\chi_{\sssty 1,-1,1,-1}(j,k,l,m) +(-1)^{\frac{d-1}{2}}\chi_{\sssty 1,1,-1,-1}(j,k,l,m) 
+\chi_{\sssty-1,1,1,1}(j,k,l,m)\nonumber \\
&+\chi_{\sssty 1,-1,1,1}(j,k,l,m)+\chi_{\sssty 1,1,-1,1}(j,k,l,m)+\chi_{\sssty 1,1,1,-1}(j,k,l,m) 
\bigg].
\label{Csum}
\end{align}
The function $ \chi _{ \eta_1,\eta_2,\eta_3,\eta_4}$ is defined by
\begin{align}
 &\chi _{ \eta_1,\eta_2,\eta_3,\eta_4}(j,k,l,m)  = |\eta_1j+\eta_2k+\eta_3l+\eta_4m|^{d-2}\\
&\nonumber\times\hspace{-4mm}\sum_{\alpha,\beta,\gamma,\delta =0}^{\frac{d-3}{2}} \frac{(\frac{d-1}{2})_\alpha(\frac{3-d}{2})_\alpha(\frac{d-1}{2})_\beta(\frac{3-d}{2})_\beta(\frac{d-1}{2})_\gamma(\frac{3-d}{2})_\gamma(\frac{d-1}{2})_\delta(\frac{3-d}{2})_\delta}{2^{\alpha+\beta+\gamma+\delta} \alpha!\beta!\gamma!\delta! (\alpha+\beta+\gamma+\delta+d-2)!}\, \frac{(\eta_1j+\eta_2k+\eta_3l+\eta_4m)^{\alpha+\beta+\gamma+\delta}}{(\eta_1j)^\alpha (\eta_2k)^\beta (\eta_3l)^\gamma (\eta_4m)^\delta}, 
 \end{align}
where $\eta_1,\eta_2,\eta_3,$ and $\eta_4$ can be either $1$ or $-1$, as in (\ref{Csum}),
and $(x)_n$ denotes the Pochhammer symbol
\begin{equation}
(x)_n = x(x+1)(x+2)...(x+n-1).
\end{equation}

The expression (\ref{Csum}) was discovered through rather unsystematic experimentation with the approximations for Jacobi polynomials of high degree in the complex plane given in \cite{Hahn}. It is not surprising that analytic simplifications occur at odd $d$. Indeed, simple closed form expressions in terms of trigonometric functions valid specifically for $d=3$ were presented in \cite{GMLL} for the AdS mode functions (\ref{modes}). Corresponding closed form expression should also exist for higher odd $d$, since there are iterative relations allowing to change $\alpha$ and $\beta$ in the Jacobi polynomials $P^{(\alpha,\beta)}_n$ by 1, and hence change $d$ in steps of 2. It is not unnatural to think that such expressions would integrate to give the interaction coefficients in a closed form with a finite number of terms, and consequently a closed expression for the asymptotics of the form (\ref{Csum}).  (Closed form expressions for the interaction coefficients at $d=3$ corresponding to gravitational nonlinearities were discussed in \cite{GMLL}.)

We do not see any obvious way to analytically connect the simple asymptotic expression (\ref{Csum}), valid at odd $d$, to the general asymptotic expression (\ref{eq:4large}), valid at any $d$. Bessel functions of half-integer order can be expressed through elementary functions, but even that does not make (\ref{Csum}) immediately obvious, except for $d=3$, where both (\ref{eq:4large}) and (\ref{Csum}) straightforwardly agree with the considerations of \cite{CEJV}.
More specifically, Bessel functions of a half-integer order can be represented as (see \cite{watson} for further details)
\beq
J_{m+1/2}(z)= \sqrt{\frac{2}{\pi z}}R_{m,1/2}(z)\sin z-\sqrt{\frac{2}{\pi z}}R_{m-1,3/2}(z)\cos z,
\eeq
where $R_{m,\nu}(z)$ are polynomials in $1/z$ known as the Lommel polynomials and given by
\beq
 R_{m,\nu}(z)=\sum\limits_{0\le n\le m/2} (-1)^n{m-n \choose n} \frac{\Gamma(\nu+m-n)}{\Gamma(\nu+n)}\left(\frac{2}{z}\right)^{m-2n}.
\eeq
Substituting these expressions in (\ref{eq:4large}) results in a finite number of fairly simple terms. However, these terms cannot be integrated individually on account of the singularities at $z=0$. (The full expression is, of course, non-singular, and the Bessel functions actually vanish at the origin.) These circumstances make recovering the explicit form of (\ref{Csum}) non-straightforward, as it requires taking into account cancellations of singularities between a large number of different terms. We have checked numerically, however, that (\ref{eq:4large}) and (\ref{Csum}) provide the same asymptotics.


\end{document}